\documentclass[iop]{emulateapj}
\usepackage{apjfonts}
\usepackage{graphicx}
\usepackage{natbib}
\usepackage{multirow}
\def\hi{H{\sc i}}

\def\deg{$^\circ$}

\def\dol{\textsc{dolphot}}

\def\n{NGC\,}

\def\apm{{$\pm$}}
\long\def\symbolfootnote[#1]#2{\begingroup%
\def\thefootnote{\fnsymbol{footnote}}\footnote[#1]{#2}\endgroup}

\newcommand{\T}{\rule{0pt}{2.8ex}}
\newcommand{\B}{\rule[-1.4ex]{0pt}{0pt}}

\shorttitle{Radial Migration in \n7793}
\shortauthors{Radburn-Smith et al.}
\submitted{Accepted for publication in ApJ}

\begin{document}

\title{Outer-Disk Populations in \n7793: Evidence for Stellar Radial Migration}
\author{David~J.~Radburn-Smith\altaffilmark{1},
  Rok~Ro\v{s}kar\altaffilmark{2},
  Victor~P.~Debattista\altaffilmark{3},
  Julianne~J.~Dalcanton\altaffilmark{1},
  David~Streich\altaffilmark{4},
  Roelof~S.~de~Jong\altaffilmark{4},
  Marija~Vlaji\'{c}\altaffilmark{4},
  Benne~W.~Holwerda\altaffilmark{5},
  Chris~W.~Purcell\altaffilmark{6},\\
  Andrew~E.~Dolphin\altaffilmark{7},
  and Daniel~B.~Zucker\altaffilmark{8,9}}

\altaffiltext{1}{Department of Astronomy, University of Washington, Seattle, WA 98195, USA}
\altaffiltext{2}{Institut f\"ur Theoretische Physik, Universit\"at
  Z\"urich, Switzerland}
\altaffiltext{3}{RCUK Fellow. Jeremiah Horrocks Institute, University of Central Lancashire, Preston, PR1 2HE, UK}
\altaffiltext{4}{Leibniz-Institut f\"ur Astrophysik Potsdam, D-14482
  Potsdam, Germany}
\altaffiltext{5}{European Space Agency, ESTEC, 2200 AG Noordwijk, The
  Netherlands}
\altaffiltext{6}{Department of Physics and Astronomy, University of
  Pittsburgh, Pittsburgh, PA 15260, USA}
\altaffiltext{7}{Raytheon, 1151 E. Hermans Road, Tucson, AZ 85756, USA}
\altaffiltext{8}{Department of Physics and Astronomy, Macquarie University, NSW 2109, Australia}
\altaffiltext{9}{Australian Astronomical Observatory, NSW 1710, Australia}

\begin{abstract}
  We analyzed the radial surface brightness profile of the spiral
  galaxy \n7793 using $HST$/ACS images from the GHOSTS survey and a
  new $HST$/WFC3 image across the disk break. We used the photometry
  of resolved stars to select distinct populations covering a wide
  range of stellar ages. We found breaks in the radial profiles of all
  stellar populations at 280\arcsec ($\sim$5.1\,kpc). Beyond this disk
  break, the profiles become steeper for younger populations. This
  same trend is seen in numerical simulations where the outer disk is
  formed almost entirely by radial migration.  We also found that the
  older stars of \n7793 extend significantly farther than the
  underlying \hi~disk. They are thus unlikely to have formed entirely
  at their current radii, unless the gas disk was substantially larger
  in the past. These observations thus provide evidence for
  substantial stellar radial migration in late-type disks.
\end{abstract}

\section{Introduction}

Studies of the chemical composition of galaxy disks typically assume
that the orbital radii of stars are essentially static. Only a small
diffusion in radial distance, increasing with age, is expected due to
dynamical heating \citep[e.g.,][]{wie77,deh98}. In this scenario, the
metallicity of stars of similar age and galactic distance should be
comparable, as they are born from the remnants of previous stellar
generations that orbited at the same radius. However, observations of
local F and G dwarfs have found little evidence for such coexistent
evolution, instead revealing a large scatter in the age--metallicity
distribution \citep{edv93}. Theoretical work has suggested that this
scatter may be due to resonant interactions between stars and spiral
arms, which enable stars to migrate across many kpc
\citep{sel02}. Indeed, using a high-resolution hydrodynamical
simulation, \cite{roskar08b} found that such migration naturally led
to the high degree of variation observed in the age--metallicity
relation of local Milky Way (MW) stars. These events, in which the
guiding center of the star shifts, differ from general diffusion where
the epicyclic motion increases by only a few kpc due to an increase in
velocity dispersion.
 
Testing such models of stellar migration may be possible using studies
of stellar populations beyond the classical disk break of late-type
galaxies. The cause of this break is still debated \citep[see the
review by][]{tru05}, however the change in surface brightness is often
associated with a threshold in the star formation rate \citep[SFR;
e.g.,][]{van81,ken89,sch04,elm06}. Using a study of 85 late-type
galaxies, \cite{poh06} found that the surface brightness profile of
approximately 60\% of disk galaxies steepens beyond the disk
break. Hence, the outskirts of disk galaxies are often associated with
a dearth of star formation. Using {\it N}-body simulations of disk
galaxy formation, \cite{roskar08} found that if star formation is
sufficiently suppressed beyond the break, then that region may form
almost entirely from stars migrated from the inner disk. This would
leave a clear signal on radial surface brightness profiles, with older
stars (i.e., those that have been subject to scattering for longer)
exhibiting shallower profiles.

\begin{figure*}
\begin{center}
\begin{tabular}{c}
\resizebox{95mm}{!}{\includegraphics{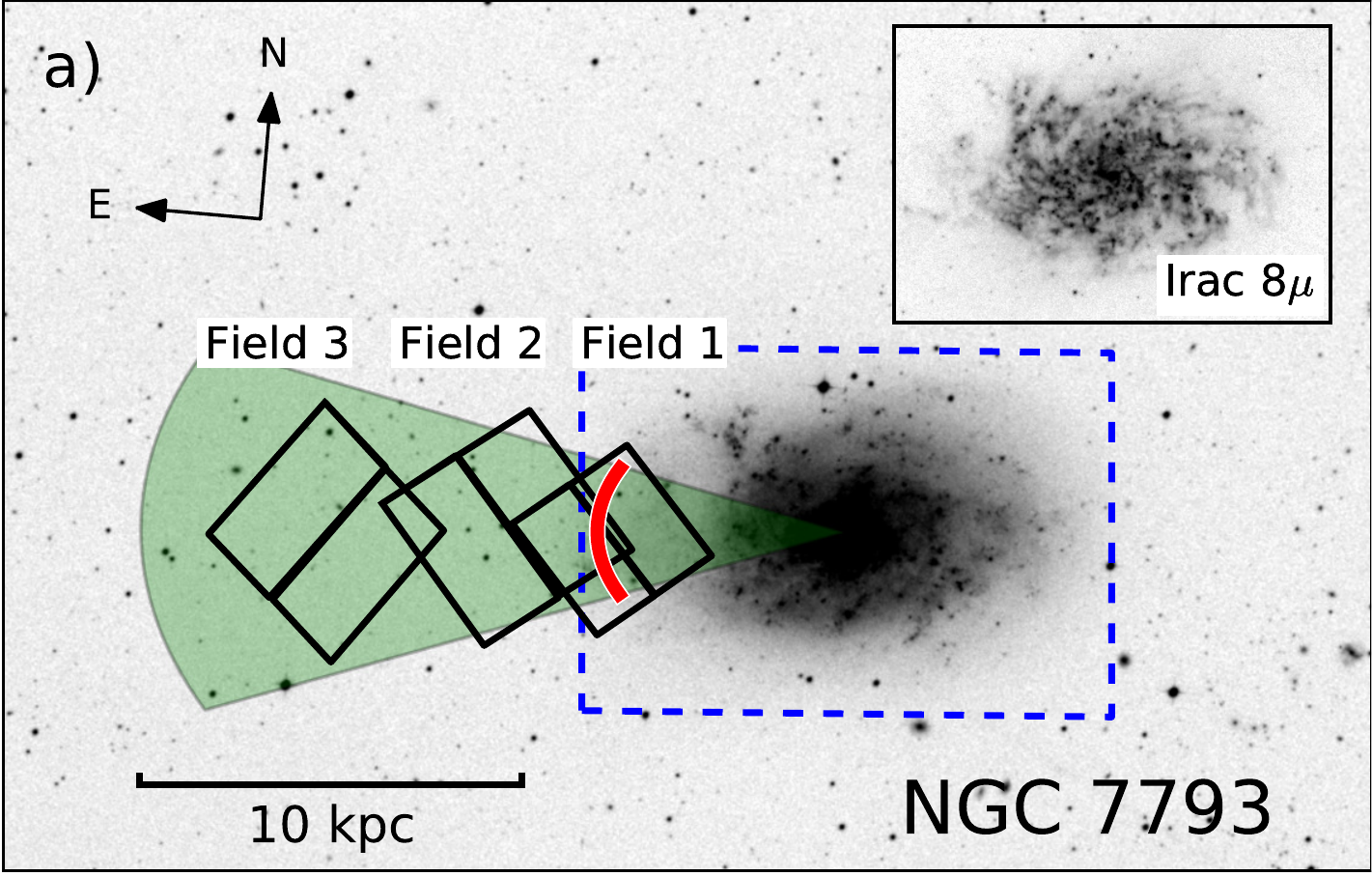}}\\
\resizebox{155mm}{!}{\includegraphics{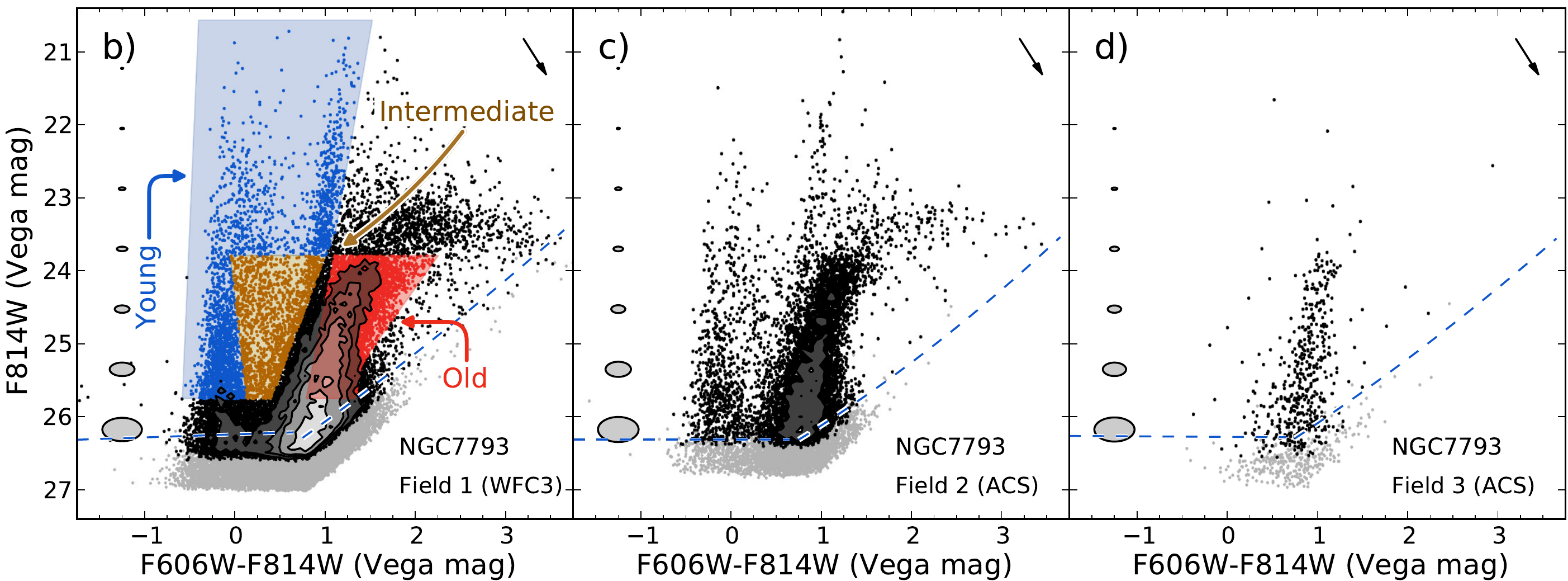}}\\
\end{tabular}
\caption{Location of the observations and resulting color magnitude
  diagrams (CMDs). Radial surface brightness profiles are taken from
  the shaded wedge (with a de-projected opening angle of 25\deg) in
  panel (a). The thick red line marks the disk break. The blue dashed
  region denotes the area covered by the \textit{Spitzer}/IRAC 8\,$\mu$m
  observation in the top right inset, which highlights the spiral
  structure that may be responsible for stellar migration. Panels
  (b)--(d) plot the individual CMDs, with the regions used to extract
  the different stellar populations, as described in the text,
  highlighted in panel (b). Gray and black points indicate different
  cuts on the signal-to-noise ratio of 3.5 and 5.0, respectively, with
  contours drawn at densities of 50, 100, 150, and 250 stars per 0.01
  mag$^2$. The black arrows indicate the direction of reddening due to
  foreground extinction, which has been removed from the data. The
  shaded ellipses indicate photometric errors as reported by \dol, and
  the blue dashed lines mark the 50\% completeness limits calculated
  from artificial star tests.}
\label{fig:cmds}
\end{center}
\end{figure*}

Observations from the GHOSTS\footnote{Galaxy Halos, Outer disks,
  Substructure, Thick disks, and Star clusters.} survey
\citep[][hereafter R-S11]{ghosts}, which have targeted the outskirts
of several nearby disk galaxies using the \textit{Hubble Space
  Telescope} Advanced Camera for Surveys ($HST$/ACS), thus offer a
significant test of radial migration. GHOSTS observations of \n4244
revealed the location of the disk break to be independent of stellar
age, thus constraining several mechanisms for the formation of this
feature \citep{dejong07}. However, lack of data beyond the break
prevented further analysis. In this paper, we focus on the spiral
galaxy \n7793. This flocculent galaxy is (1) relatively nearby
(3.7\apm0.1\,Mpc, R-S11), (2) situated at high Galactic latitude
(--77.2\deg) and hence relatively uncontaminated by foreground stars,
and (3) an isolated loosely bound member of the Sculptor Group. \n7793
was recently studied by \cite{vlajic2011}, who found no sign of a
break in the red giant branch (RGB) population, but did find evidence
for an upturn in the radial metallicity distribution beyond the disk
break.  In this paper, we expand on this work by using data from the
GHOSTS survey together with a new $HST$ Wide Field Camera 3 (WFC3)
image of the inner disk to measure fainter stars and to recover
stellar densities in this crowded region. We compare these
observations with predictions from the radial migration simulations of
\cite{roskar08} and discuss alternate explanations.

\section{$HST$ ACS/WFC3 Photometry}
\label{sec:data}

Figure~\ref{fig:cmds} plots the location of the two GHOSTS fields
discussed in this paper (Fields 2 and 3). Each field was observed with
dithered 740\,s exposures in both the F606W and F814W filters. R-S11
described the data reduction and photometry of the resolved stars in
these fields using the \dol~code \citep{dol00}. A further field (Field
1) was observed with the WFC3 camera to characterize the disk
truncation ({\it{HST}} GO program 12196). This observation was made using the
same filters as the ACS fields, with dithered exposures of 980\,s in
F606W and 1488\,s in F814W. We use the same \dol~parameters and the same
crowded-field stellar selection criteria for the WFC3 field as used by
R-S11 for the ACS photometry. Despite minor differences between the
cameras, namely, pixel scale and quantum efficiency in the F814W
filter, we found little discrepancy in the resulting photometry
between the cameras in overlapping regions.

The resulting color magnitude diagrams (CMDs) for each of the fields
are presented in Figure~\ref{fig:cmds}. By analyzing synthetic CMDs
generated with a constant SFR of 0.01\,$M_{\sun}$ yr$^{-1}$, an
appropriate metallicity of [Fe/H]$=-1$, and photometric errors
matching the observations, we identified three regions of the CMD
corresponding to three discrete age ranges. The area marked as
``Young'' consists of bright main sequence and helium-burning (HeB)
stars, which are collectively 10--100\,Myr old; the ``Intermediate''
region is composed of $100\textrm{\,Myr\,}$--1\,Gyr old HeB stars; and
the ``Old'' region consists of $1-10$\,Gyr old RGB stars. We
subsequently use these regions, which suffer little cross
contamination, to select stellar populations by age.

R-S11 generated artificial stars for the ACS fields, which were
subsequently used to assess errors and correct for completeness due to
stellar crowding. We generated an additional 1.8 million artificial
stars for the WFC3 image. Figure~\ref{fig:cmds} plots the 50\%
recovery limit of these artificial stars, which we set as our limiting
magnitude.

\begin{figure*}
\begin{center}
\resizebox{180mm}{!}{\includegraphics{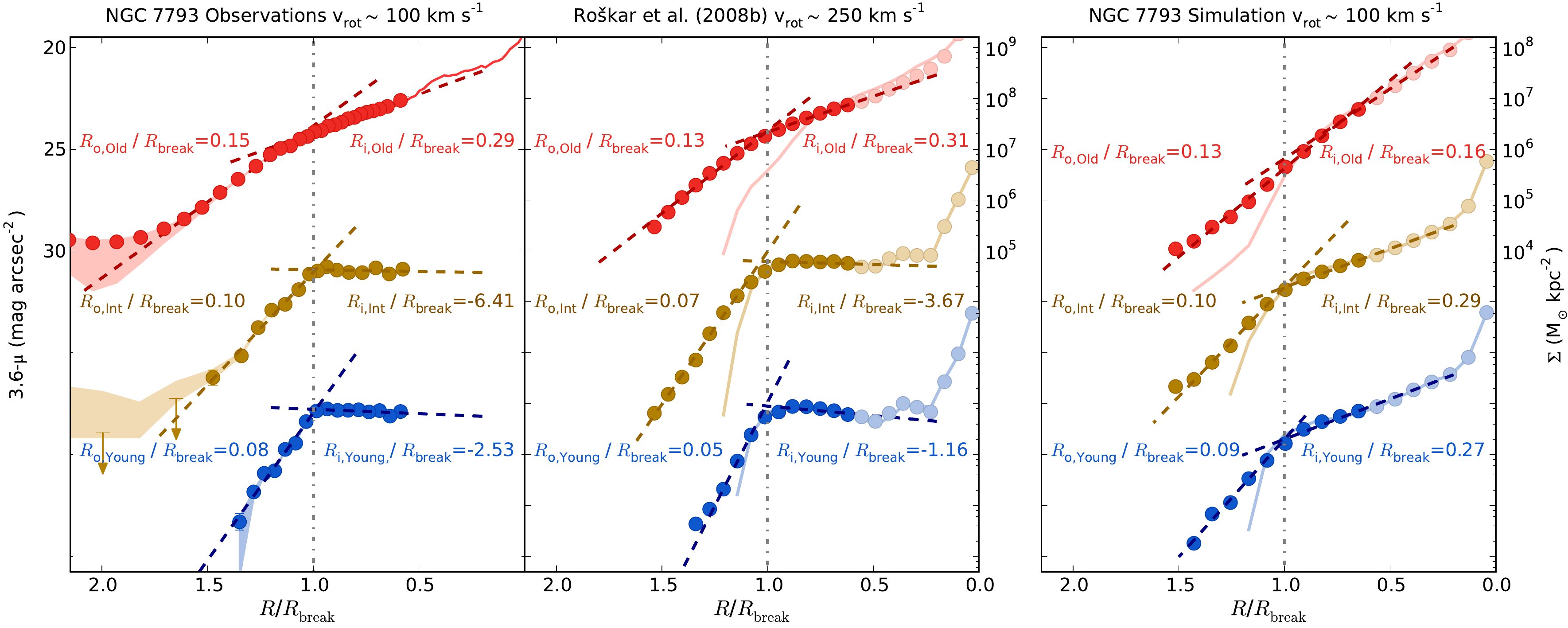}}
\caption{Radial surface brightness profiles of ``Old'' (red),
  ``Intermediate'' aged (yellow), and ``Young'' (blue) stellar
  populations. The left panel plots number counts in \n7793 taken from
  the shaded region of Figure~\ref{fig:cmds}, panel (a). The middle
  and right panels, respectively, plot surface densities extracted
  from similar regions of the \cite{roskar08} simulation of a Milky
  Way (MW) sized galaxy and the new simulation of \n7793 described in
  the text. The \n7793 observations have been linearly offset such
  that the ``Old'' population matches a {\it{Spitzer}} 3.6\,$\mu$m
  surface brightness profile, indicated by a solid red line. For
  clarity, the ``Intermediate'' and ``Young'' population are
  arbitrarily offset by 5 and 12.5\,mag, respectively, in the
  observations, and by 2 and 4 dex in the simulations. All profiles
  are scaled to the approximate location of the ``Young'' break
  (280\arcsec in \n7793, 10.2\,kpc in the MW simulation, and 7.7\,kpc
  in the \n7793 simulation). Arrows indicate where the Poisson
  uncertainty may move the data point below the background
  level. Dashed lines indicate exponential fits to the inner and outer
  profiles, with the corresponding normalized exponential-disk scale
  lengths given by $R_i$ and $R_o$, respectively. The shaded regions
  in the left panel indicate the range of possible profiles due to
  different treatments of background contamination as described in the
  text. Faint solid lines in the simulation panels plot the radial
  surface brightness profiles using the formation radius of the stars,
  i.e., removing the effects of stellar migration.}
\label{fig:radial}
\end{center}
\end{figure*}

We found good agreement between detection rates and photometry in the
overlapping region of Fields 1 and 2. The WFC3 observation from Field
1 yielded 3474 detections brighter than F814W = 25.75\,mag after
culling, while ACS (Field 2) found 3631, despite the shorter exposure
time. Approximately 82\% of the ACS detections were coincident between
fields. After correcting for completeness with the artificial star
tests, this increased to 95\%. The remainder can be attributed to
incompleteness at the magnitude limit of the survey. The median
photometric differences between coincident detections are
\mbox{$-0.05$} and 0.03\,mag in F606W and F814W, respectively. The
mean error of these detections, when combined in quadrature, is
$\pm$0.10 in both filters. Hence, although the offsets are systematic,
they are significantly less than the \dol~uncertainties. We thus found
our photometry from the two different cameras to be consistent and
complete to $\textrm{F814W}=25.75$\,mag.

\section{Radial Surface Brightness Profiles}
\label{sec:profiles}

In the left panel of Figure~\ref{fig:radial}, we plot the radial star
counts of \n7793 within the shaded region of panel (a) of
Figure~\ref{fig:cmds} for our key stellar populations. These profiles
are all normalized to the radial break observed in the youngest
population. In addition to masking foreground stars and background
galaxies, a region of significant star formation, located at
$23^{\mathrm{h}}58^{\mathrm{m}}13$\fs1, $-32^{\circ}36'$37\farcs3 was
masked out of Fields 1 and 2 to provide a better estimate of the mean
surface brightness. We assigned an equivalent surface brightness to
the star count profiles by multiplying the logarithm of the older RGB
counts by 2.5 and linearly offsetting the result to match a
\textit{Spitzer} Infrared Array Camera (IRAC) 3.6\,$\mu$m surface
brightness profile extracted from the same region. Consequently, we
find the star count profiles reach a limiting magnitude equivalent to
$\sim32$\,AB mag arcsec$^{-2}$ at 3.6\,$\mu$m.

Each of the stellar population bins denoted in panel (b) of
Figure~\ref{fig:cmds} are substantially brighter than the completeness
limits. They are thus less susceptible to spurious detections from
cosmic rays and CCD noise that become significant at a lower
signal-to-noise ratio. Instead, contamination is expected from
unresolved background galaxies, which far outnumber any foreground MW
stars. To account for this, we followed the same analysis used by
R-S11 and ran the photometry code on ACS images of high-$z$ sources
with exposure lengths longer than our observations. As these fields
are free of any resolvable stars outside of the MW, any detections
will thus be contaminants, primarily from background galaxies. For
each of the denoted stellar populations in panel (b) of
Figure~\ref{fig:cmds}, we measured contamination from the number
density of remaining detections in these ``empty'' archival fields
after imposing the selection criteria used by R-S11. For the ``Old'',
``Intermediate'', and ``Young'' bins we recovered $4.4\times10^{-4}$,
$4.0\times10^{-4}$, and $3.3\times10^{-4}$ detections per
arcsec$^{2}$, respectively. Each of these contaminant measurements are
lower than any of the detection rates in the corresponding \n7793
populations. Averaging the outermost radial bins of each population in
\n7793, we, respectively, measured $4.0\times10^{-3}$,
$4.5\times10^{-4}$, and $1.6\times10^{-3}$ detections per
arcsec$^{2}$. As discussed in \cite{ell07}, these can be treated as
upper limits on the true level of contaminants in each bin. Moreover,
using the \textsc{GalaxyCount} software described in \cite{ell07}, we
can infer uncertainties on these values due to cosmic variance and
Poisson noise. These equate to upper limits on the total number of
background galaxies in an ACS/WFC field of $163\pm 14$, $18\pm 4$, and
$65\pm 8$ for the respective age bins. These uncertainties are
incorporated into the final error we assign to the profiles. The lower
bound on the profiles due to removing this higher level of
contamination, as well as an upper bound resulting from not
subtracting any contaminants, are indicated by the shaded regions in
the left panel of Figure~\ref{fig:radial}.

We are able to probe the outskirts of \n7793 much deeper than previous
integrated light studies, whose profiles do not cover these radii or
reach this surface brightness limit
\citep[e.g.,][]{mun09,vlajic2011}. We consequently found that the disk
break, $R_{\rm{break}}$, occurs at approximately the same location
regardless of the age of the underlying stars, in agreement with
observations of the edge-on disk galaxy \n4244 \citep{dejong07}. This
break occurs at 280\arcsec (5.1\,kpc), internal to the regions covered
by \cite{vlajic2011} and external to the integrated light profiles of
\cite{mun09}. \cite{dejong07} argued that a constant break location
calls for a dynamical interpretation, given that a break radius due
solely to a star formation threshold is unlikely to remain at the same
radius over the lifetime of the galaxy, due to disk growth. We note
that the break is weakest in the oldest population, which also shows a
possible second break to a shallower profile around $\sim1.8\times
R_{\rm{break}}$. This may correspond to a transition in the number
counts from a disk-dominated population to one composed primarily of
stellar halo stars (see Section~\ref{sec:thickdisk}).

Although the disk break appears at a constant radius for all stellar
ages, the profiles in Figure~\ref{fig:radial} suggest that the slope
of the profile does depend on age, both within and beyond the break.
We characterize this slope by fitting the inner and outer profiles to
an exponential surface brightness function of the form
$\mu(r)=\mu_0e^{-{R}/{R_d}}$, where $\mu_0$ is the central surface
brightness and $R_d$ is the exponential-disk scale length.  The
resulting fits, shown in Figure~\ref{fig:radial} and
Table~\ref{table:fits}, reveal a smooth increase in the steepness of
the outer profile with decreasing age, together with a corresponding
flattening of the inner profile.

\begin{deluxetable}{lr@{$\pm$}lr@{$\pm$}lr@{$\pm$}l}
\tablecaption{Exponential Outer-disk Fits\label{tab:disks}}
\tablehead{
{\rule{0pt}{2.ex}}Data Set &
\multicolumn{2}{c}{$R_\textrm{old}/R_\textrm{break}$} &
\multicolumn{2}{c}{$R_\textrm{int}/R_\textrm{break}$} &
\multicolumn{2}{c}{$R_\textrm{young}/R_\textrm{break}$} 
}
\startdata
\multicolumn{7}{c}{\n7793 Observations}\\
\hline
\T Raw            & 0.147 & 0.007 & 0.107 & 0.016 & 0.078 & 0.014\\
Corrected         & 0.145 & 0.007 & 0.098 & 0.016 & 0.075 & 0.014\\
\B Overcorrected & 0.134 & 0.016 & 0.080 & 0.017 & 0.063 & 0.022\\
\hline
\multicolumn{7}{c}{\rule{0pt}{2.3ex}Simulations}\\
\hline
\T\cite{roskar08} & 0.134 & 0.008 & 0.072 & 0.007 & 0.053 & 0.006\\
\B\n7793 Model    & 0.125 & 0.012 & 0.101 & 0.007 & 0.094 & 0.007\\
\hline
\multicolumn{7}{c}{\rule{0pt}{2.3ex}Simulations without Migration}\\
\hline
\T\cite{roskar08} & 0.025 & 0.018 & 0.018 & 0.011 & 0.024 & 0.014\\
\n7793 Model      & 0.054 & 0.019 & 0.032 & 0.011 & 0.025 & 0.013
\enddata 
\tablecomments{The ``Raw'' \n7793 observations refers to profiles
  uncorrected for contamination from background galaxies. For the
  ``Overcorrected'' profiles, an upper limit on the number of the
  contaminants is inferred from the average density of detections in
  the outermost radial bins of each stellar population in the raw
  data. The exponential-disk fits to the ``Simulations without
  Migration'' use the formation radii of stars born in the outer-disk
  region.}
\label{table:fits}
\end{deluxetable}

To assess the uncertainty in the profiles we use Monte Carlo
realizations of the data. As summarized in Table~\ref{table:fits}, the
outer-disk parameters of the ``Old'' population are robust ($\sim$5\%
uncertainty), but low number counts in the ``Intermediate'' and
``Young'' outer-disk profiles leads to uncertainties of $\sim$16\% and
$\sim$19\%, respectively. Accounting for different levels of
contamination has little effect on the fits.

\section{Comparison with Simulations}
\label{sec:migration}

In isolated simulations of a MW-type galaxy, \cite{roskar08} found
significant migration of stars over several kpc, such that $\sim$90\%
of the outer disk was predominantly formed from migrated stars. Older
populations exhibited larger scale lengths than younger populations,
because the former were subject to migration for longer. We can
compare our measured profiles to those of \cite{roskar08}, which are
plotted in the middle panel of Figure~\ref{fig:radial} for the same
age ranges used in our CMD analysis. These profiles were extracted
from the simulation at the same inclination as \n7793 ($\sim$53\deg),
and for comparison with the observations are similarly scaled to the
break radius of the youngest stellar population. The behavior of the
profiles in the simulation is qualitatively similar to what is
observed in the data. However, the simulation is not necessarily the
best analog for \n7793. The mass of the simulation contained within
10\,kpc is approximately $1.2\times10^{11} M_{\sun}$, larger than the
mass of \n7793, which \cite{car90} measure within 7.35\,kpc as
$1.5\times10^{10}\,M_{\sun}$. Similarly, the radius of the disk break
is also larger in the simulation (10.2\,kpc cf. 5.1\,kpc in \n7793).

\subsection{Modeling NGC 7793}

A potentially better analog to \n7793 is the equivalent simulation
generated to model \n300 in \cite{gog10}. This galaxy is another
member of the Sculptor Group with mass similar to \n7793, although the
star formation history (SFH) of the system and the lack of a disk
break indicate a quiescent formation. \cite{gog10} found little
evidence of migration in $HST$ observations of \n300, and less
migration in the \n300 simulation than the \cite{roskar08} model. This
led the authors to suggest that both mass and environment may be
factors in radial migration. As the formation of transient spiral
structure is linked to gas accretion and star formation \citep{sel84},
such a dependence on environment is to be expected. This was further
demonstrated by \cite{gog10}, who note that the comparably massive
M\,33, which is interacting with M\,31, does show signs of radial
diffusion \citep{wil09}.

\begin{figure*}
\begin{center}
\resizebox{160mm}{!}{\includegraphics{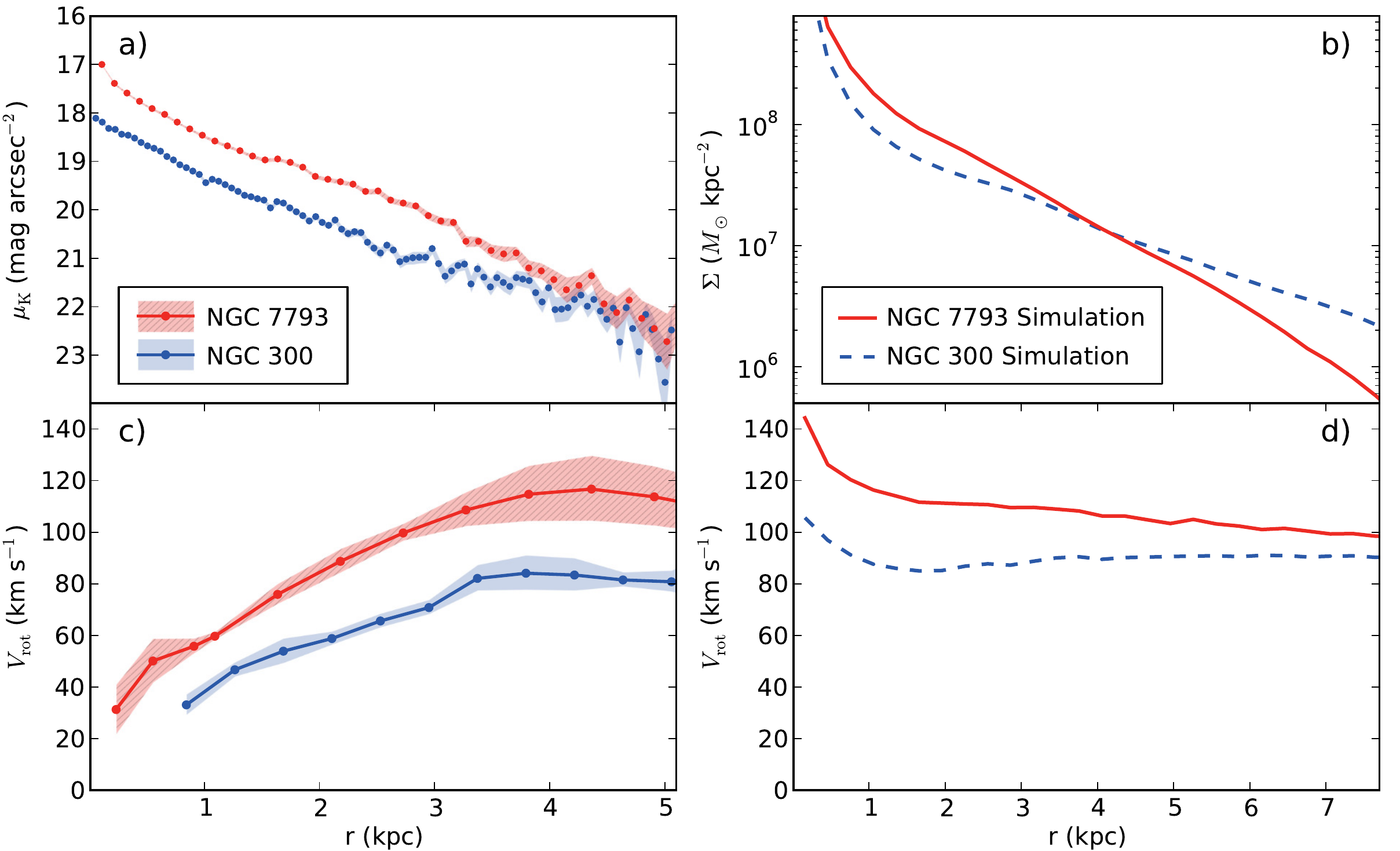}}
\caption{Radial surface brightness profiles (top panels) and
  \hi~rotational velocity profiles (bottom panels) from observations
  of \n7793 and \n300 (left panels) and from simulations of \n7793 and
  \n300 (right panels). The \n7793 simulation is described in the text
  while the \n300 simulation is taken from \cite{gog10}. The $K$-band
  surface brightnesses plotted in panel (a) were measured by
  \cite{mun07} from the Two Micron All Sky Survey Extended Source
  Catalog \citep{jar00}. The shaded regions indicate the 1$\sigma$
  uncertainty in the photometry from variations in intensity within
  each isophote and errors in sky subtraction. The \hi~rotation curves
  were measured by \cite{car90} and \cite{puc90} from Very Large Array
  observations of \n7793 and \n300, respectively. The shaded regions
  indicate the velocity uncertainties derived from the largest
  difference between the average radial rotation and the rotation
  curve measured on each side of the galaxy. Data are shown to the
  disk break in \n7793 for both the observations ($\sim5.1$\,kpc) and
  simulations ($\sim7.7$\,kpc).}
\label{fig:comparison}
\end{center}
\end{figure*}

We extend this analysis by observing that the $K$-band absolute
surface brightness of \n7793 is on average $\sim$1.0\,mag
arcsec$^{-2}$ brighter than \n300 (see panel (a) of
Figure~\ref{fig:comparison}). The surface density of the \n300 disk
may thus be too low to support the transient spiral structures
required to drive substantial migration. To better match the
properties of \n7793 we resimulated the \n300 model with a smaller
spin parameter to yield a disk with a smaller scale length
(Figure~\ref{fig:comparison}, panel (b)), larger rotational velocity
(Figure~\ref{fig:comparison}, panel (d)), and higher surface density
within the central 4\,kpc. These changes resulted in a more
concentrated system, with $1.7\times10^{10}\,M_{\sun}$ contained
within the disk break (located at 7.7\,kpc). The stellar disk in this
model was also subject to significant radial migration, leading to a
similar increase in disk scale length with stellar age as in the
\cite{roskar08} model (see right panel of Figure~\ref{fig:radial} and
Table~\ref{table:fits}). The effect is, however, weaker. We partly
attribute this difference to a relative increase in the level of in
situ formation beyond the disk break in the \n7793 model. This is
manifest in Figure~\ref{fig:radial} and Table~\ref{table:fits} by
exponential-disk fits to the formation radii of stars located beyond
the break. Specifically, the normalized scale lengths are greater in
the \n7793 model than in the MW simulation. The \n300 model generated
by \cite{gog10} shows a further increase in this in situ population,
which resulted in an even softer disk break.

\begin{figure}
\begin{center}
\resizebox{82mm}{!}{\includegraphics{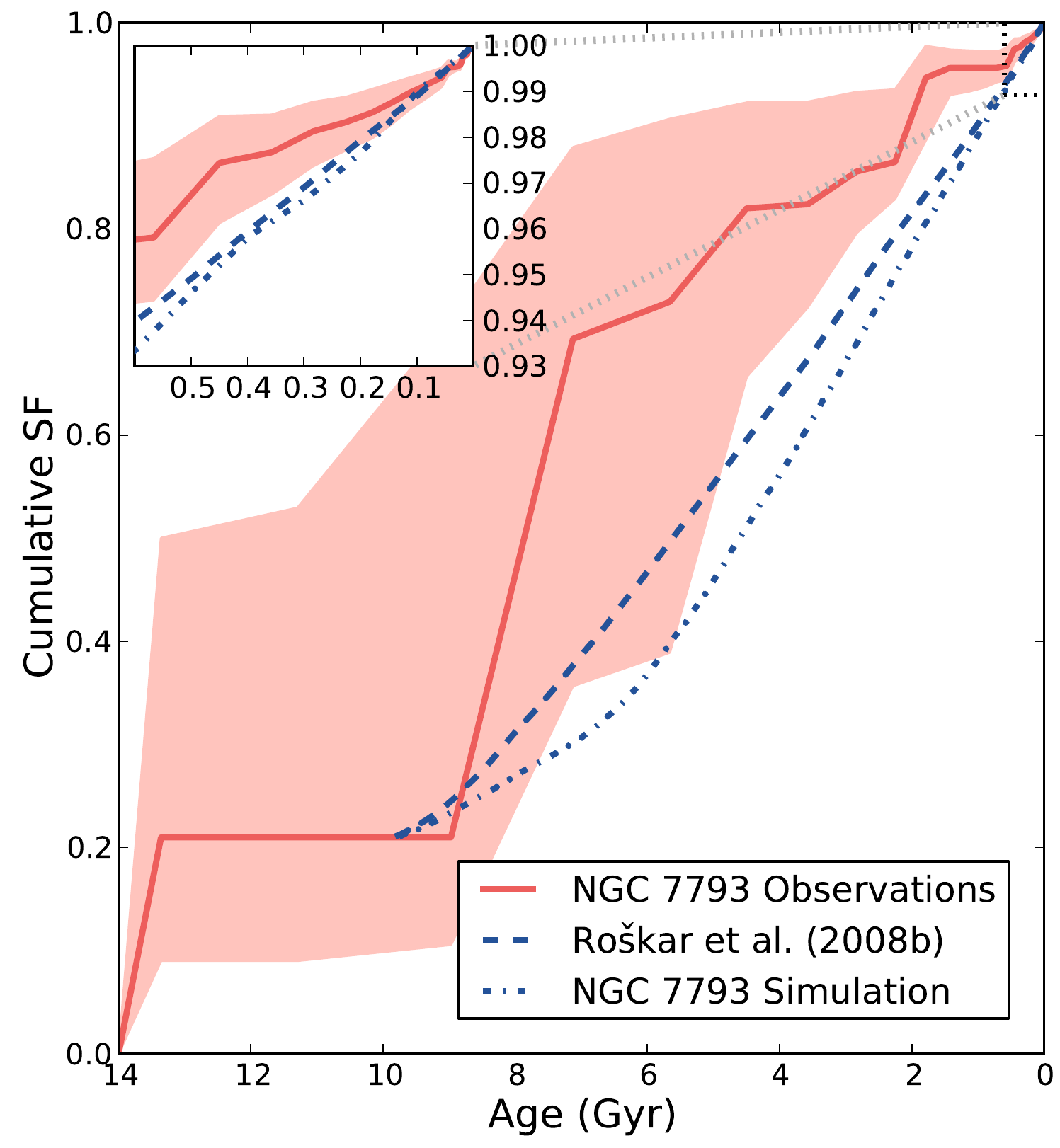}}
\caption{Comparison of the cumulative star formation history of \n7793
  with the MW and \n7793 simulations as a function of age. The SFH for
  \n7793 was generated using {\textsc{match}}, with the shaded region
  denoting the uncertainty in the measurement.  As the simulations are
  evolved for only 10\,Gyr, we fix their initial cumulative star
  formation to that of \n7793 10\,Gyr ago. A zoom-in of the last
  600\,Myr is shown as an inset. The isolated simulations show a
  smooth buildup of stellar mass over the lifetime of the systems,
  which is consistent with the SFH reconstructed from observations of
  \n7793.}
\label{fig:sfh}
\end{center}
\end{figure}

Figure~\ref{fig:sfh} compares the cumulative SFH of the main disk of
\n7793 to the \cite{roskar08} MW simulation and the new \n7793
simulation, both of which show signs of radial migration. To extract
the SFH from the $HST$/ACS photometry, we used the software package
{\textsc{match}} \citep{dol02}. This program uses a maximum likelihood
analysis to fit the observed CMD with stars generated from the stellar
evolution models of \citet[][with updated AGB models from
\citealp{mar08} and \citealp{gir10}, and assuming a \citealp{sal55}
IMF and a binary fraction of 40\%]{gir02}. The artificial CMD was
calculated using a 36$\times$25 grid in $\log(\rm{age})$ -- $[Z]$
space with bin widths of 0.1\,dex. We left the distance modulus and
extinction free to be fitted by {\textsc{match}}, which converged on a
distance modulus of $m-M=27.83$ and an extinction of $A_V=0.05$. These
are consistent with the literature values of $m-M=27.87\pm0.08$
(R-S11) and $A_V=0.06$ \citep{sch98}. We constrained the chemical
enrichment history to an increasing metallicity.

The overall uncertainty in the SFH, denoted in Figure~\ref{fig:sfh} by
the shaded region, is a combination of both random and systematic
uncertainties. The former are measured from the 1$\sigma$ spread in
300 Monte Carlo realizations of the best-fit solution. The systematic
uncertainties in the isochrones are calculated by varying the
$M_{\rm{bol}}$ and $T_{\rm{eff}}$ values of the underlying models
\citep[for details, see][]{dol12}.

As the simulations are only evolved for 10\,Gyr, we tie their initial
cumulative star formation to the equivalent value measured in \n7793
10 Gyr ago. The resulting SFHs of the models show a relatively smooth
build up of mass. This is to be expected, as the simulations are not
set in a full cosmological environment.  The SFH of \n7793, as
reconstructed from the observations, is consistent with a similar
steady SFR for the life of the system.

\begin{figure}
\begin{center}
\resizebox{86mm}{!}{\includegraphics{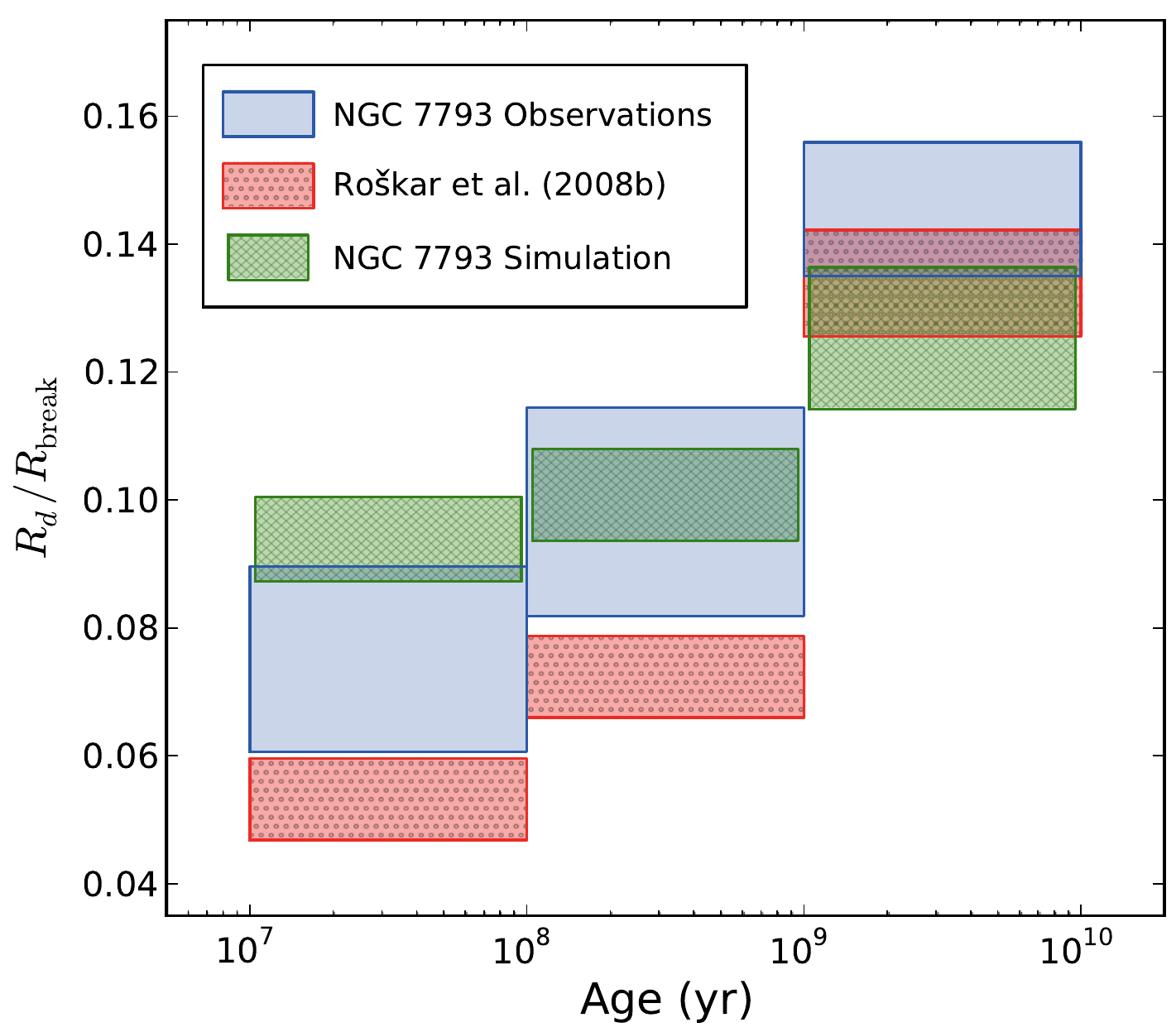}}
\caption{Comparison of the normalized exponential-disk scale lengths
  measured in Figure~\ref{fig:radial} for the outer disk of \n7793,
  the \cite{roskar08} MW simulation, and the new \n7793
  simulation. The width of each box represents the range of ages
  covered by each population, while the height encodes the 1$\sigma$
  uncertainty in the exponential-disk fit as calculated from Monte
  Carlo resamples.  The width of the \n7793 simulation boxes are
  slightly reduced for clarity.  Both observations and simulations
  argue for a smooth increase in the scale length of the outer-disk
  profile with stellar age.}
\label{fig:gradients}
\end{center}
\end{figure}

Even though the SFHs of the two simulations are similar, they exhibit
markedly different inner-disk profiles (see Figure~\ref{fig:radial}).
This suggests dynamical processes dominate the construction of the
disk profiles, whether internally by secular processes \citep[as
in][]{roskar08} or by externally driven merger processes related to
the formation of the host halo \citep[e.g., the two controlled
simulations presented by][]{pur11}. Outer-disk profiles are likely
similarly affected, and in the case of the simulations presented here,
they predominantly arise from radial
diffusion. Figure~\ref{fig:gradients} presents a direct comparison of
these normalized outer-disk gradients as a function of age. There is
remarkable agreement between the simulations and the \n7793
observations, which thus strongly attests to significant migration in
\n7793.

\subsection{Alternate Simulations}

\cite{sanchez09} have also addressed radial migration in their
cosmological simulations. They argue that the upturn in the age
profile beyond the disk break is principally due to an extended and
constant SFR in the outer disk. However, they do find evidence for
some migration, and moreover have studied the outer regions of their
cosmological simulations without the effects of such migration (see
their Figure~17). Specifically, they found a break still formed in the
radial surface brightness profiles, but without migration, the
decrease in gradient of the outer-disk profile with increasing stellar
age was lost.

Stellar radial migration may also be driven by other mechanisms, such
as resonances between spiral structures and central
bars\footnote{Although we note that resonances with the central bar
  are likely to have less influence on the outer disk
  \citep{deb06,min12}.}  \citep{min10}, or interactions with
satellites \citep{qui09,bir12}. These processes may operate in
addition to the resonant scattering from transient spiral arms, and so
serve to amplify the effect.  However, as \n7793 is not host to a
significant bar, and as such features are believed to be robust
\citep{she04,ath05,deb06}, the disk is unlikely to have been affected
by a central bar. Conversely, as detailed in Section~\ref{sec:harass},
\n7793 may have undergone a recent merger. Such an event could enhance
spiral structure in the disk and so increase the level of stellar
migration.

\section{Alternate Explanations}
\label{sec:alternate}

The increase in the outer-disk scale length with stellar age, as seen
in Figure~\ref{fig:gradients}, corresponds to an increase in the
average age of the outer disk with radius. This trend, although not
ubiquitous, has been confirmed using spectroscopic stellar ages in
several nearby disk galaxies \citep{yoa12}. To assess whether radial
migration is the only viable mechanism for setting this trend, we
consider alternate explanations below.

\subsection{Thick Disk and Halo Contamination}
\label{sec:thickdisk}
The thick disk typically extends farther radially than the thin disk
by a factor of 1.25, and in low-mass galaxies can account for nearly
half the total mass and luminosity \citep{yoachim06,com11}. As the
thick and thin disks can only be separated in high inclination
systems, a significant portion of the older outer-disk stars may
actually lie in an extended thick disk component. The smooth change in
gradient of the outer radial profile with age argues against a simple
two component model. However, if true, the coincident break radius of
both old and young populations suggests that the mechanism propagating
the break, and thus setting the outer-disk gradients, applies equally
to both components. Furthermore, the thick disk itself may have formed
from a migrated population \citep{sch09,loe10}. Hence, contamination
from a thick disk is unlikely to affect the inference of migration.

In the ``Old'' RGB population of \n7793 plotted in
Figure~\ref{fig:radial}, a gradual departure from the outer
exponential is seen below 29\,mag arcsec$^{-2}$. This may indicate a
transition to either a thick disk or old stellar halo component. At
$r=11$\,kpc ($2.2\times R_{\mathrm{break}}$), this extra component
accounts for some $4.5\times10^{-3}$\,stars\,arcsec$^{-2}$. This is
approximately a factor of 10 greater than the number of predicted
contaminants, and similar to the surface densities attributed to the
stellar halos of isolated simulations of disk--satellite interactions
from \cite{pur10}. However, such low numbers will have minimal impact
on the exponential fits summarized in Figure~\ref{fig:gradients}.

\subsection{In Situ Formation}
\label{sec:insitu}

Star formation in outer disks has been characterized by
\cite{bigiel10a,bigiel10b}, who find evidence for spatial correlation
between far-ultraviolet (FUV) emission, which traces recent star
formation, and emission from atomic hydrogen gas (\hi), which
comprises most of the interstellar medium in the outer disk. Although
\cite{bigiel10b} find the overall star formation efficiency to be
extremely low in these regions, they observe that this efficiency
scales with \hi~gas density. Hence, as the gas disk of \n7793 extends
beyond the disk break, we expect some of the stars in this region to
have formed in situ. This may partly account for the vertical offset
seen in Figure~\ref{fig:gradients} between the \n7793 observations and
the \cite{roskar08} simulation.

\begin{figure}
\begin{center}
\resizebox{87mm}{!}{\includegraphics{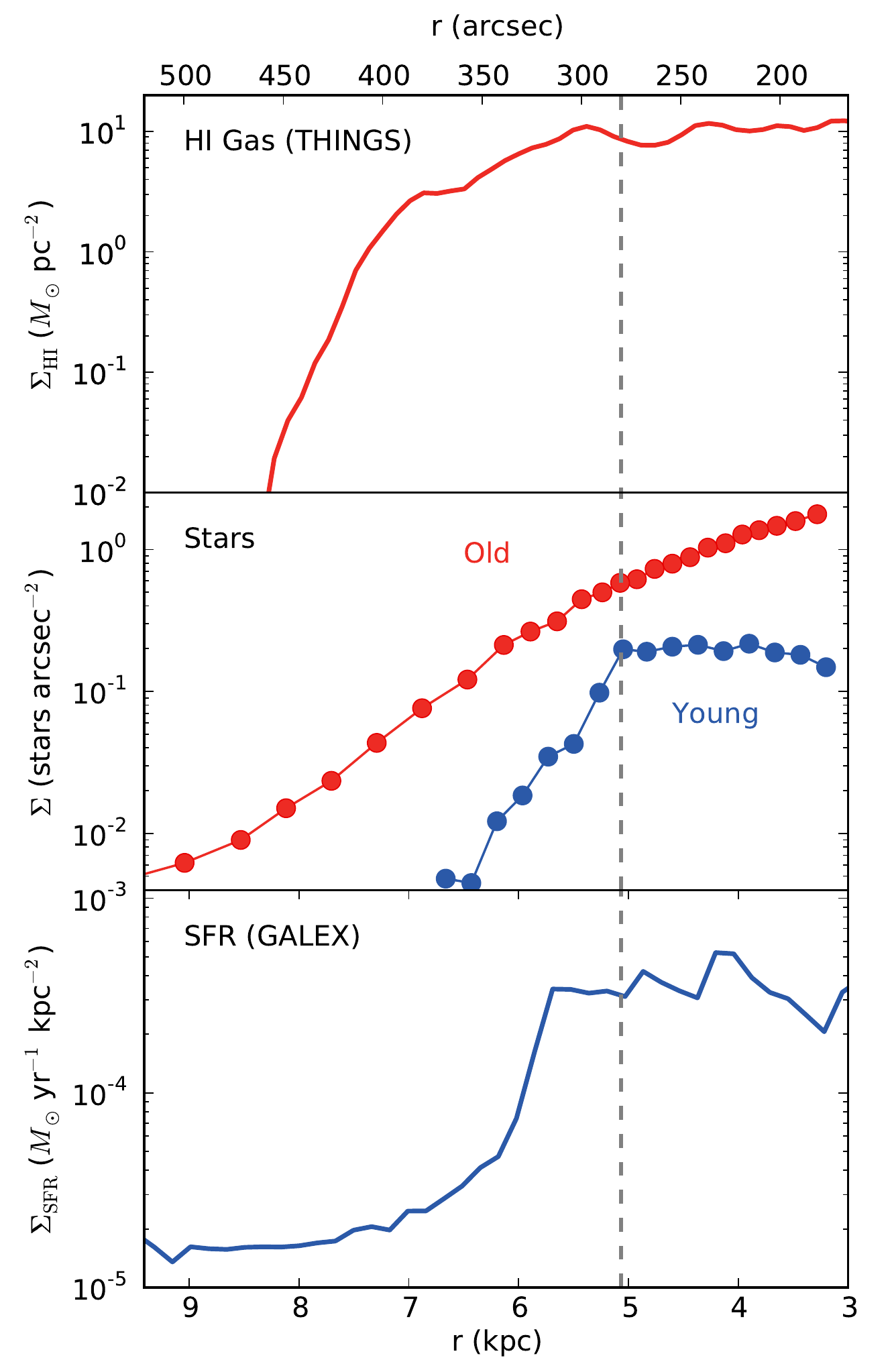}}
\caption{Top: the \hi~gas surface density in \n7793 from the THINGS
  survey. Middle: the star counts corresponding to the old RGB (red)
  and young MS and HeB (blue) profiles shown in
  Figure~\ref{fig:radial}. Bottom: the star formation rate density
  from {\it{GALEX}} FUV emission. The old RGB stars are found to
  extend much farther than the \hi~disk.}
\label{fig:profiles}
\end{center}
\end{figure}

To explore this possibility, Figure~\ref{fig:profiles} plots the
surface densities of \hi~gas mass from the THINGS survey
\citep{walter08}, and the SFR inferred from the {\it{GALEX}}
Ultraviolet Atlas of Nearby Galaxies \citep{gil07}. The FUV emission
does indeed extend slightly beyond the truncation. However, using
{\textsc{match}} with this inferred SFR to generate artificial CMDs
underestimates the number of ``Young'' stars observed by almost an
order of magnitude (although this calculation does carry significant
uncertainties). The stars may instead have formed in previous episodes
of formation not probed by the FUV flux. Indeed, the gas disk extends
further still, although, as noted by \cite{vlajic2011}, the \hi~disk
of \n7793 is unusually small. As shown in Figure~\ref{fig:profiles},
the radial extent of the gas is significantly smaller than the RGB
distribution. Thus, the bulk of the old RGB stars were not likely
formed in situ from the \hi~gas reservoir now present in the region,
unless the \hi~disk was substantially larger in the past.

\subsection{Accretion/Harassment}
\label{sec:harass}

Many interacting systems are evident in the Local Group. Our own
galaxy is tidally disrupting the Small and Large Magellanic Clouds,
producing a number of phenomena such as the Magellanic Stream
\citep{mat74,bru05}. Such interactions strongly affect the
distribution of material in outer disks
\citep{you07,san08,qui09,bir12,kaz08}. In the optical, \n7793 does not
appear to be undergoing such a major merger, which would likely
disrupt the outer-disk profiles. However, the \hi~disk of this
relatively isolated galaxy is warped, and studies have revealed
non-circular motions on the north side, indicative of harassment
\citep{car90}, or of an advanced stage of merging
\citep[e.g.,][]{san99}. Such minor interactions may explain the lack
of gas in the outskirts, and may lead to an extended and heated
component \citep{pur10,kaz09}. Moreover, such interactions are also
likely to induce the transient spiral features that drive stellar
radial migration \citep{pur09,pur11,str11}. As these transient spiral
arms can last for a couple of Gyr \citep[and potentially much longer
in star-forming galaxies due to swing amplification processes
involving giant molecular clouds, e.g.,][]{don12}, the level of
stellar radial migration we infer in \n7793 may be a direct
consequence of a completed past merger.

\section{Summary}
\label{sec:summary}

We have measured the radial surface brightness profiles of three
distinct stellar populations across both the inner and outer disk of
\n7793. Breaks in the radial profiles are found at the same location
regardless of age. However, older stars show a steeper profile
internal to the break and a shallower profile beyond the break in
comparison to younger stars. The observed smooth increase in the
outer-disk scale length with age suggests a formation mechanism that
continually operates over a significant fraction of the lifetime of
the galaxy.

We have compared the surface brightness profiles to {\it{N}}-body
simulations that exhibit significant radial stellar diffusion,
including a new simulation of a system comparable in mass to
\n7793. The gradients of the outer-disk profiles in these simulations
are set almost entirely by migration due to resonant scattering. The
new simulation shows a slightly higher level of in situ formation
beyond the disk break relative to \cite{roskar08}, which corresponds to
a flattening of the profiles. However, the gradients overall match
extremely well to the outer disk of \n7793.

We thus argue that the observations presented here are indicative of
high levels of stellar radial migration in \n7793. This inference
carries significant ramifications for studies of stellar populations
in disk galaxies, including the MW.

Support for this work was provided by NASA through grants GO-10889 and
GO-12196 from the Space Telescope Science Institute, which is operated
by the Association of Universities for Research in Astronomy,
Incorporated, under NASA contract NAS5-26555.

\end{document}